\documentclass[iop,apj,twocolumn,a4paper]{aastex631}
\usepackage{amsmath,amstext}
\usepackage{hyperref}
\usepackage{amssymb}
\usepackage{wasysym}
\usepackage{float}
\usepackage{ulem}
\usepackage{breakurl}
\usepackage{color}

\def\msun{\,{M_\odot}}

\def\spose#1{\hbox to 0pt{#1\hss}}
\def\lta{\mathrel{\spose{\lower 3pt\hbox{$\mathchar"218$}}
     \raise 2.0pt\hbox{$\mathchar"13C$}}}
\def\gta{\mathrel{\spose{\lower 3pt\hbox{$\mathchar"218$}}
     \raise 2.0pt\hbox{$\mathchar"13E$}}}

\def\tMS{t_{\rm MS}}
\def\tHZ{t_{\rm HZ}}

\lefthead{Madau}
\righthead{Beyond the Drake Equation}
\journalinfo{To appear in The Astrophysical Journal}


\begin{document}

\title{Beyond the Drake Equation: A Time-Dependent Inventory of 
Habitable Planets and Life-Bearing Worlds in the Solar Neighborhood}

\author[0000-0002-6336-3293]{Piero Madau}
\affiliation{Department of Astronomy \& Astrophysics, University of California, 1156 High Street, Santa Cruz, CA 95064}
\affiliation{Dipartimento di Fisica ``G. Occhialini", Università degli Studi di Milano-Bicocca, Piazza della Scienza 3, I-20126 Milano, Italy}

\begin{abstract}
We introduce a mathematical framework for statistical exoplanet population and astrobiology studies that may help direct future observational efforts and experiments. The approach is based on a set of differential equations and provides a time-dependent mapping between star formation, metal enrichment, and the occurrence of exoplanets and potentially life-harboring worlds over the chemo-population history of the solar neighborhood.  Our results are summarized as follows: (1) the formation of exoplanets in the solar vicinity was episodic, starting with the emergence of the thick disk about 11 Gyr ago; (2) within 100 pc from the Sun, there are as many as $11,000\,(\eta_\oplus/0.24)$ Earth-size planets in the habitable zone (``temperate terrestrial planets" or TTPs) of K-type stars. The solar system is younger than the median TTP, and was created in a star formation surge that peaked 5.5 Gyr ago and was triggered by an external agent; (3) the metallicity modulation of the giant planet occurrence rate results in a later typical formation time, with TTPs outnumbering giant planets at early times; and (4) the closest, life-harboring Earth-like planet would be 
$\lta 20$ pc away if microbial life arose as
soon as it did on Earth in $\gta 1$\% of the TTPs around K stars. If simple life is abundant (fast abiogenesis), it is also old, as it would have emerged more than 8 Gyr ago in about one-third of all life-bearing planets today. Older
Earth analogs are more likely to have developed sufficiently complex life capable of altering their environment and producing detectable oxygenic biosignatures.
\bigskip
\end{abstract}

\keywords{Astrobiology -- Exoplanets -- Habitable Planets -- Metallicity -- Solar Neighborhood  -- Star Formation}

\section{Introduction}

The search for habitable exoplanets and extraterrestrial life beyond the solar system is a topic of central interest for modern science, and one of the  most compelling and consequential endeavors for humankind. 
Simple life emerged on Earth within the first billion years of its habitable window, and the high frequency of terrestrial planets in the habitable zones (HZs) around GK dwarf stars inferred from NASA's {Kepler} observations \citep[][]{Bryson2021} invites the question of how often (if at all) life may have arisen on other worlds in the past. This pursuit will ultimately require statistical analyses of the population of habitable systems, in-depth studies of the climates of individual planets, and searches for chemical biomarkers \citep{Schwieterman2018}, and has motivated the development of the next generation of large ground-based facilities and instrumentation. 
The yield and characterization of Earth-like planets will be a primary science metric for future space-based flagship missions, but the optimal observational strategy for addressing the origin  and properties of planetary systems and the prevalence of habitable exoplanets and life beyond the solar system remains unclear \citep[e.g.,][]{Bean2017,Tasker2017,Sandora2020,Truitt2020,Checlair2021,Sarkar2022,Batalha2023}. The gathering of comprehensive data for each individual system is impractical if not impossible, so a statistical perspective is necessary to prioritize targets for follow-up observations. In particular, one would like to identify -- given a model of habitability and biosignature genesis -- how potential biosignature yields change during the evolution of a stellar system as a function of stellar properties like age, mass, and metallicity.

This paper aims to present a theoretical framework for exoplanet population and astrobiology studies that may provide a better statistical understanding of the formation history, frequency, age, and metallicity distributions of different planet types around stars of different properties. Our approach may also  establish a useful basis for testing hypotheses about habitable environments and life beyond the solar system,
for gaining a sense of biosignature yields, and for informing future observational efforts and experiments. A well-known parameterization of the present-day abundance of life-bearing worlds in the Galaxy, $N_\ell$, is represented by the first four terms in the probabilistic Drake equation \citep{Drake1965}, which can be rewritten as 
\begin{equation}
N_\ell=N_{\rm MS}(t_0)\,f_p\,n_e\,f_\ell.
\label{eq:Drake}
\end{equation}
Here, $N_{\rm MS}(t_0)$ is the total number of stars that are on the main sequence today and can provide their planets a stable HZ, $f_p$ is the fraction of these stars that have planetary systems, 
$n_e$ is the average number per planetary system of Earth-size planets that are in the HZ, and $f_\ell$ is the subset of these rocky exoplanets that are ``Earth-like" in a more detailed biochemical and geophysical sense and where simple life eventually arises. The first three  terms ($N_{\rm MS}, f_p, and n_e$)
in Equation (\ref{eq:Drake}) have already experimental measurements, and the fourth ($f_\ell$) is a conditional probability that may potentially be observable in the coming decades via  spectroscopic searches for biosignature gases in exoplanet atmospheres \citep{Schwieterman2018,Seager2018}. In Drake's famous formulation, in order to estimate the number of active, communicative extraterrestrial civilizations around us today, the right-hand side of Equation (\ref{eq:Drake}) gets multiplied by the fraction $f_i$ of life-bearing planets on which intelligent life emerges, times the percentage $f_c$ of such civilizations that produces a detectable signal, times the fractional longevity $f_L$ of a technological species. 

The Drake equation and its `biosignature' version \citep{Seager2018} 
amount to a  pedagogical and organizational summary of the factors that may affect the likelihood of detecting technologically advanced civilizations or just simple microbial life evolving on a habitable planet. They are only meant to guide the observational inputs needed to make an educated guess, rather than provide a time-dependent mapping between star formation, environment, exoplanets, and life-harboring worlds. Their inherent limitations include both the lack of temporal structure \citep[e.g.,][]{Cirkovic2004,Forgan2009,Cai2021,Kipping_note2021} --
an assumption of uniformity with time that precludes the inclusion of evolutionary effects associated with, e.g., the star formation and chemical enrichment history of the local Galactic disk and the time line of life emergence -- as well as the difficulty of casting in a probabilistic argument the variety of phenomena and associated timescales that may influence anything quantified by probability $f$-factors and multiplicity $n_e$.
Recent rapid developments in astrophysics and planetary sciences warrant a more informative and modern evolutionary framework, a {\it rate-equation approach} based on a system of first-order differential equations. These describe the changing rates of star, metal, planet, and habitable  world formation over the history of a given stellar system, and can easily be adapted to incorporate the hierarchy of astrophysical and biological processes that regulate the age-dependent inventory of any key planet population. 

The field of Galactic habitability and the formation history of Earth-like and giant planets in the Milky Way and the Universe as a whole have been research topics for more than two decades \citep[e.g.,][]{Gonzalez2001,Lineweaver2004, Gowanlock2011,Behroozi2015,Gobat2016,Zackrisson2016,Forgan2017,Balbi2020}.
Our approach expands upon some of the ideas employed in these early papers and develops new ones, focusing instead on the time-varying incidence of exoplanets and potentially habitable worlds over the chemo-population history of the {\it solar neighborhood}, the target of current and next-generation stellar and planetary surveys. This is the locale where more detailed calculations are justified by an avalanche of new data and actually needed in order to estimate, given a model of habitability and biosignature genesis, the relative biosignature yields among potential target stars. 

The plan of this paper is as follows. In Section~2 we present the basic rationale and main ingredients of our modeling: the star formation history (SFH) and metallicity distribution function (MDF) in the solar vicinity, the planet occurrence rate around GK stars, and various metallicity-dependent effects. 
In Section~3 we cast and integrate our rate equations for the time evolution of the local abundance of dwarf stars and the giant planets and rocky planets in the HZ around them. We track giant planets to gauge the impact of their enhanced occurrence rate at higher host star metallicities relative to the weaker frequency-metallicity correlation of terrestrial planets. 
In Section~4 we extend our formalism to speculate about the formation history of life-harboring environments in the local volume under the hypothesis of a rapid abiogenesis process on Earth-like planets, and estimate the prevalence of nearby biospheres in terms of the exoplanet census as a whole. Finally, we summarize our findings and conclusions in Section~5.

\section{Basic Stellar and Planetary Astrophysics}

In order to provide absolute number counts in the  solar vicinity we shall use the recent tally
of main-sequence stars, giants, and white dwarfs within 100 pc of the Sun from the {Gaia Early Data Release 3} (EDR3). The {Gaia} Gatalog of Nearby Stars contains 
\begin{equation}
N_\star(t_0)=331,312
\end{equation}
objects and is estimated to be $>92$\% complete down to faint stellar type M9 \citep{Smart2021}.
Apart for a minor correction (a fraction of a percent for the initial mass function
(IMF) in Equation
(\ref{eq:Kroupa})) associated with the contribution of remnant neutron stars and black holes, $N_\star(t_0)$ represents the total number of stars ever formed in the solar neighborhood.
Below, we shall use this normalization together with an SFH and an IMF to compute the number of main-sequence stars as a function of time.

\subsection{Star Formation History}

\begin{figure}[t]
\centering
\vspace{0.cm}
\includegraphics[width=\hsize]{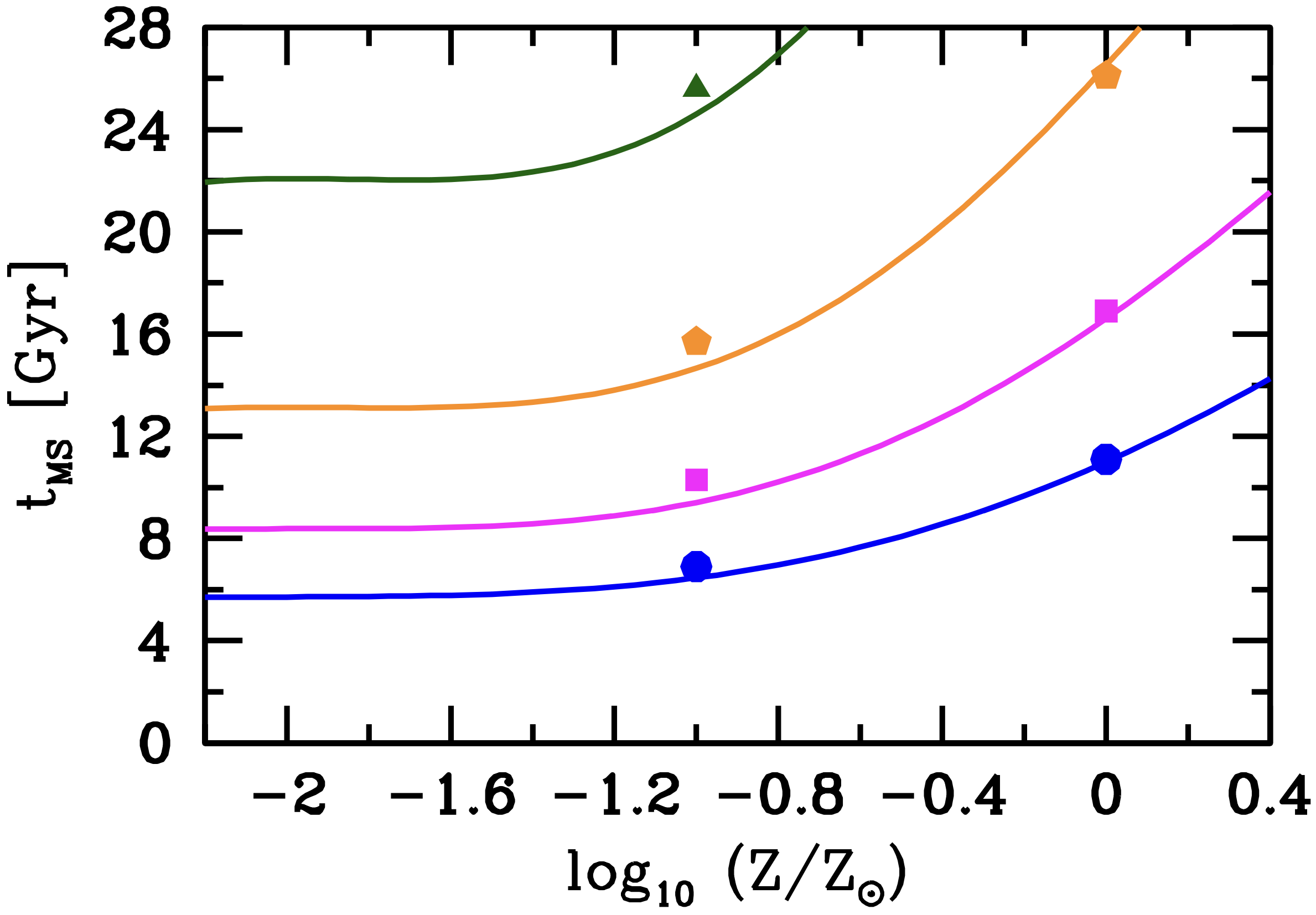}
\caption{Main-sequence lifetimes 
$t_{\rm ms}$ for stars with masses of 0.7, 0.8, 0.9, and 1.0 $\msun$ (from top to bottom) at different metallicities, $-2.2<\log_{10} (Z/Z_\odot)< 0.4$
(\citealt{Hurley2000}, ``old" solar composition). The points show the results of the more recent stellar evolution calculations by \citet[][the ``enhanced oxygen abundance" model]{Truitt2015}. 
}
\label{fig:tms}
\end{figure}

Let $\phi(m)$ and $\psi(t)$ be the (universal) IMF and SFH by number, respectively. The IMF and SFH are normalized so that $\int_0^{t_0} \psi(t)dt=1= \int_{m_l}^{m_u} \phi(m)dm$. The number of stars that are on the main sequence at time $t$, $N_{\rm MS}(t)$,
evolves at the rate
\begin{equation}
\dot N_{\rm MS}(t) = N_\star(t_0) 
\int \phi(m) [\psi(t) - \psi(t-\tMS)]dm,
\end{equation}
where the dot denotes the time derivative, $\tMS(m)<t$ is the main-sequence lifetime, and the second term in the square brackets corrects the rate of newly formed main-sequence stars for the number of stars that have evolved off the main sequence. In the following, we shall assume a \citet{Kroupa2001} IMF
\begin{equation}
\phi(m) = 
\begin{cases} 
0.2530\,m^{-1.3} &~~~(0.08\le m<0.5)\\
0.1265\,m^{-2.3} &~~~(0.5\le m<100),
\end{cases}
\label{eq:Kroupa}
\end{equation}
where $m$ is measured in solar masses. The main-sequence lifetime $\tMS$ can be computed using the analytical fitting formulae  of  \citet{Hurley2000} (based on the evolutionary tracks of \citealt{Pols1998})
as a function of $m$ and  metallicity $Z$ (see Fig. \ref{fig:tms}).\footnote{Other stellar evolution models could be adopted \citep{Valle2014,Truitt2015,Stancliffe2016}, but the resulting changes would not be significant in this context, as uncertainties in the input physics and solar composition lead to errors that are small
compared to those associated with, e.g., uncertainties in the local SFH and exoplanet occurrence rates.}\,
A G-type $m=1$ main-sequence star, for example, has a lifetime of $\tMS=11\,$Gyr at solar metallicity, and  $\tMS=6.5\,$Gyr at $Z=0.1\,Z_\odot$.

The SFH of the solar neighborhood has been recently reconstructed  by \citet{Alzate2021} using 
all 120,452 stars brighter than $G=15$ mag  within 100 pc of the the Sun in the {Gaia} DR2 catalog. In broad agreement with previous determinations based on different techniques and data sets \citep[e.g.,][]{Snaith2015,Mor2019,Ruiz-Lara2020}, their results show two main early episodes of star 
formation: (1) a peak of activity occuring 10 Gyr ago that produced a significant number of stars with sub-solar metallicities, followed by a star formation minimum (quenching) around 8 Gyr ago; and (2) a more recent burst about 5.5 Gyr ago. Since then, star formation has been declining until recent times, making stars with supersolar metallicities in short-lived
bursts of activity.

Most low-metallicity $10$ Gyr old stars belong to the thick Galactic disk population \citep[e.g.][]{Robin2014,Haywood2019}, as opposed to the thin disk for the rest of them. Clear evidence of a thick-disk peak at age $9.8\pm 0.3$ Gyr is also seen in the local white dwarf luminosity function by \citet{Fantin2019}. An intense phase of star formation between 9 and 13 Gyr ago during the emergence of the thick disk, producing about as much mass in stars as that manufactured in the next 8 Gyr, and followed by a minimum in the star formation rate at an age of $\sim 8\,$ Gyr, was also apparent in the SFH reconstruction of \citet{Snaith2015,Snaith2014}. In a 2 kpc bubble around the Sun, \citet{Ruiz-Lara2020} inferred three recent episodes of enhanced star formation dated $\sim$ 5.7, 1.9 and 1 Gyr, in synchrony with the estimated Sagittarius dwarf galaxy pericenter passages.

\begin{figure}[!h]
\centering
\vspace{0.cm}
\includegraphics[width=\hsize]{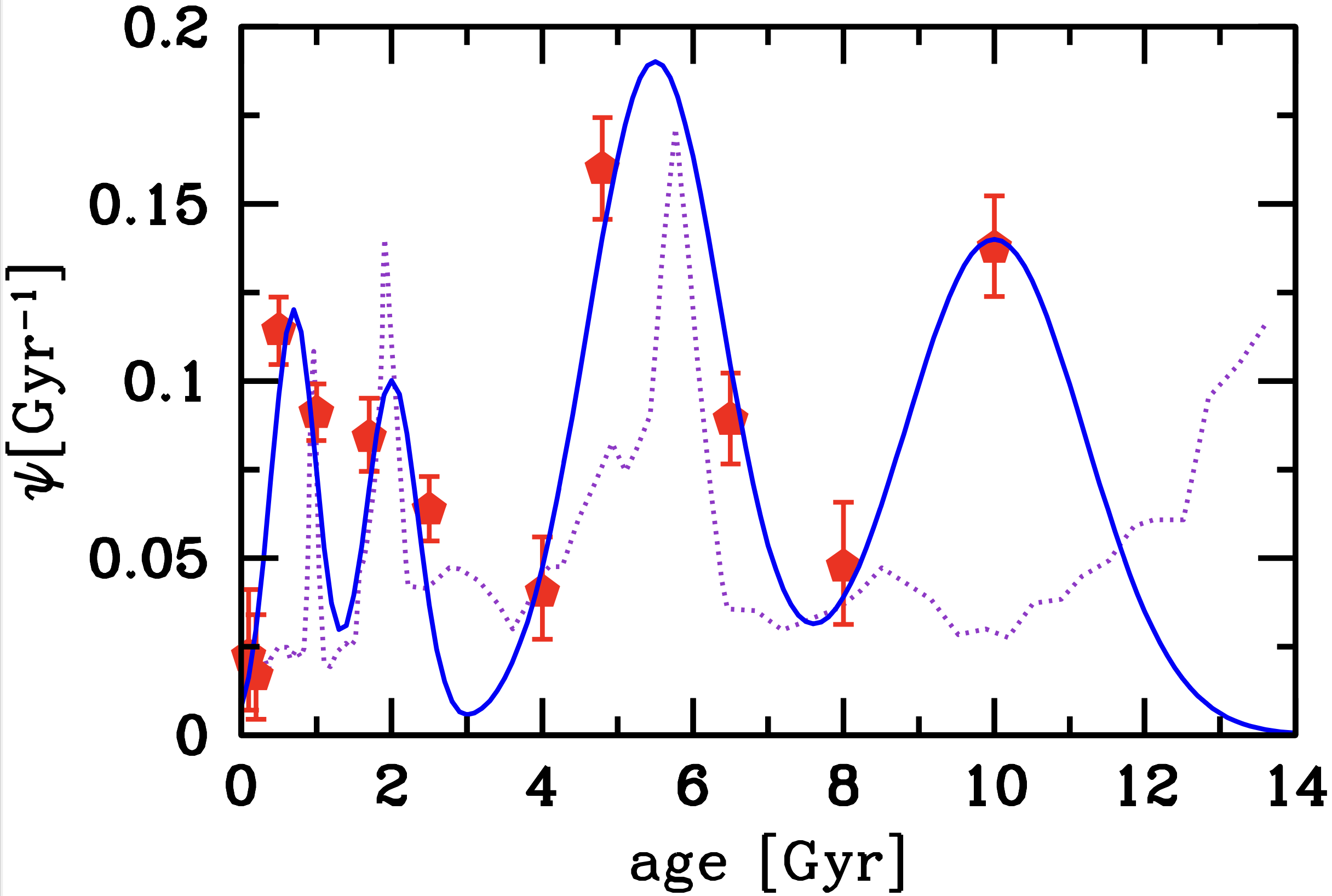}
\caption{The SFH of the solar neighborhood. The blue pentagons with error bars show the marginalized age distirbution (the fraction of stars formed per unit time at age $t_0-t$) inferred by \citet[][grid C, extinction corrected]{Alzate2021} for stars brighter than $G=15$ mag 
%within 100 pc of the Sun 
in Gaia DR2. The solid curve displays a reasonable reconstruction of the local SFH involving four Gaussians centered at ages of $10, 5.5, 2.0$ and 0.7 Gyr, having widths of 1.2, 0.9, 0.35, and 0.3 Gyr, respectively, and relative peaks 1:1.36:0.71:0.86. The dotted line indicates the three late peaks in the SFH (with a different normalization for illustration purposes) of the (kinematically defined) thin stellar disk derived by \citet{Ruiz-Lara2020}. 
}    
\label{fig:SFH}
\end{figure}

Figure \ref{fig:SFH} shows the marginalized posterior SFH of solar neighborhood stars from \citet{Alzate2021} together with 
a reasonable reconstruction involving four Gaussians centered at  ages of $10, 5.5, 2.0$ and 0.7 Gyr, and having widths of 1.2, 0.9, 0.35, and 0.3 Gyr, respectively. In this reconstruction, which we  use in the rest of this paper, about one-third of all stars belong to the old $>9\,$ Gyr population, and 
only 17\% to the youngest, $<3\,$ Gyr component. There are periods of very little star formation around $7-8$ Gyr ago and then again 3 Gyr ago. Note that, although the  analysis by \citet{Alzate2021} uses a local sample, radial migration predicts that stars in the close solar vicinity may represent a  mixture of stars born at various Galactocentric distances over the disk \citep[see, e.g.,][and references therein]{Lian2022}.

\begin{figure}
\centering
\vspace{+0.5cm}
\includegraphics[width=\hsize]{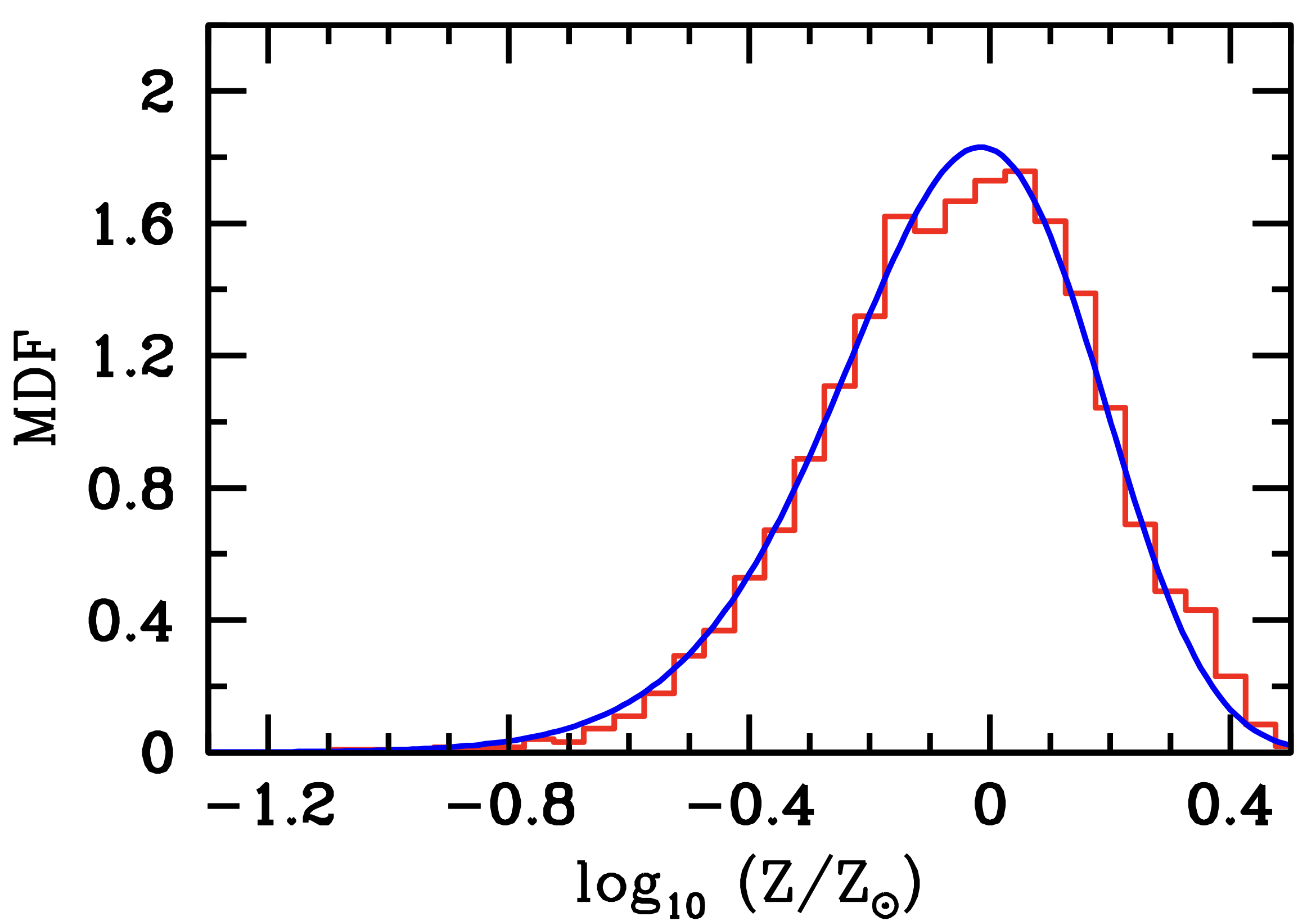}
\caption{MDT of the GALAH$+$TGAS sample (red histogram; \citealt{Buder2019}). The distribution is skewed toward the metal-poor tail, with 59\% of the stars having metallicities below solar. Note that, for plotting purposes, the histogram is expressed in densities and not in frequencies. The blue line shows a best-fit distribution in bins of $M$ of the form $G(M)= a(10^M)^b\,\exp(-c 10^M)$, where $M\equiv \log_{10} Z/Z_\odot$, $a=128$, $b=4.1$, and $c=4.25$. The mean metallicity, standard deviation, and skewness of the distribution are provided in the text.
}    
\label{fig:MDF}
\end{figure}

\subsection{Metallicity PDF and 
Age-Metallicity Relation}

To study the influence of stellar metallicity on the occurrence rates of planets and planetary systems, we shall adopt here the MDF from the GALAH$+$TGAS spectroscopic survey of dwarf stars in the solar galactic zone \citep{Buder2019}. Figure \ref{fig:MDF} shows the data histogram and the corresponding best-fit skewed  distribution, $G(M)$, with moments $\langle M\rangle=-0.07$, $\sigma_M=0.23$, and  Skew $=-0.51$, that accounts for the asymmetry of the extended metal-poor tail as well as the sharper truncation of the MDF on the metal-rich side. Here and below, we use the symbol $M$ interchangeably with $\log_{10} (Z/Z_\odot)$ for compactness, 
and the metallicity distribution $g(Z)$ is related to the MDF in bins of M as $g(Z)=G(M)/(10^M\ln10)$. 

\begin{figure}[h!]
\centering
\vspace{+0.cm}
\includegraphics[width=\hsize,
trim={0.05cm  0  0 0},clip]{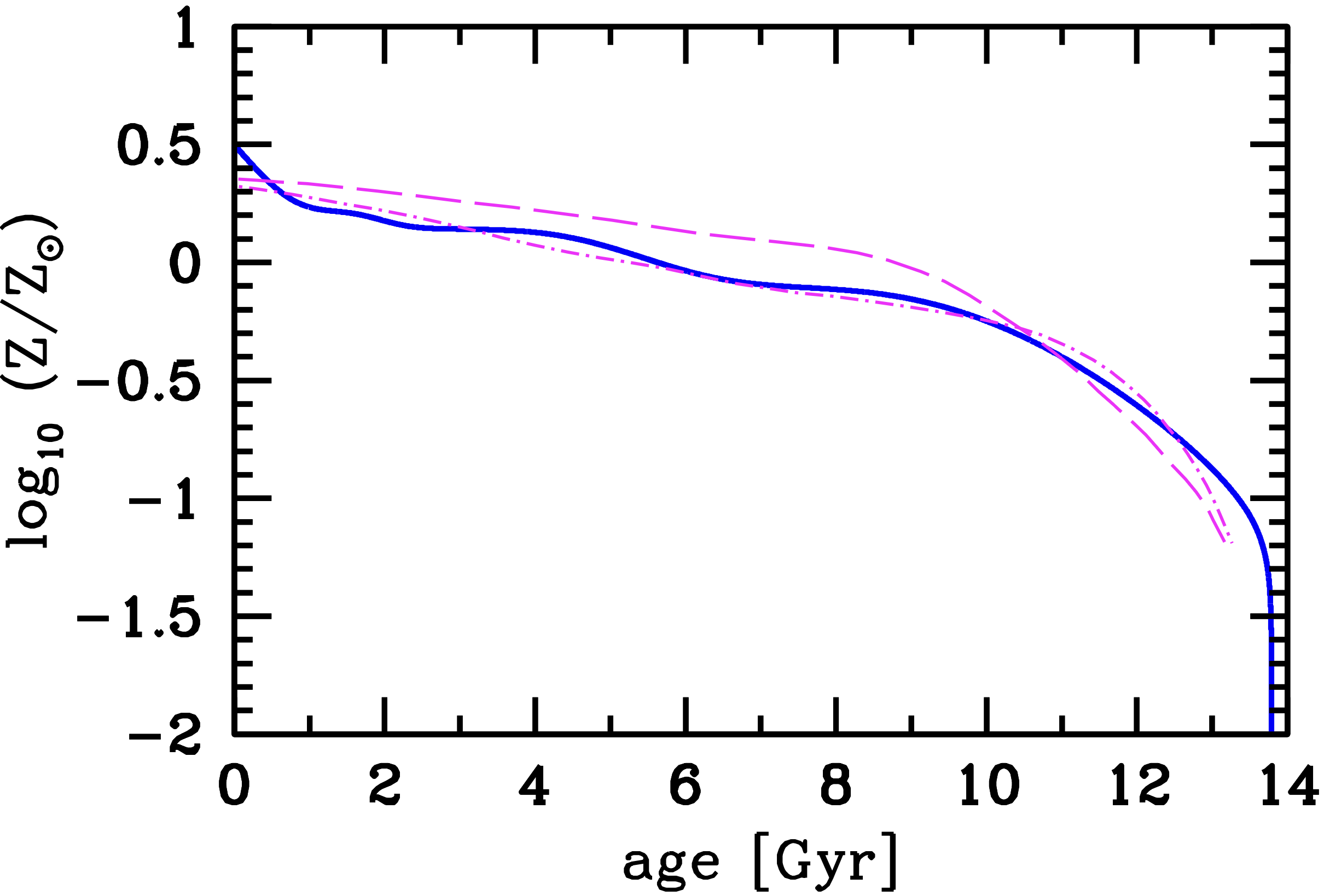}
\caption{Stellar age-metallicity relation in the solar neighborhood. The solid curve shows the result of the integration of Eq. (\ref{eq:Zdot})
for the assumed SFH $\psi(t)$ (Figure \ref{fig:SFH}) and MDF $g(Z)$ (Figure \ref{fig:MDF}). The inferred relationship is in broad agreement with the mean abundance trends vs. age recovered by \citet{Snaith2015} for the inner (dashed curve) and outer 
(dotted-dashed curve) Milky Way disk.
Note that stars in the solar vicinity have none of the characteristics of inner disk stars, and are better described as outer disk objects
\citep{Haywood2019}.
}    
\label{fig:Z}
\end{figure}

Solar neighborhood stars exhibit an age–metallicity relation, such that young age correlates with high metallicity, 
a temporal sequence that is the fossil record of the enrichment history of the Galactic disk \citep[e.g.,][]{Haywood2019,Hayden2015,Haywood2013}.
Observations have shown that this relation has a significant scatter, attributed to the effects of radial migration and chemical mixing. Since our aim here is to characterize only the prevalent metallicity in each star formation episode correctly, we shall impose an age-metallicity relationship  ignoring  the dispersion around the mean -- this is found to increase steadily with stellar age from 0.17 dex at age 2 Gyr to 0.35 dex at 13 Gyr \citep{Buder2019}. 
Within this framework, the fraction of stars that formed between time $t$ and $t+dt$, $\psi(t)dt$, is then equal to the fraction of stars with metallicity between $Z(t)$ and $Z(t)+dZ$, $g(Z)dZ$. The typical stellar metallicity therefore evolves with time as 

\begin{equation}
\dot Z(t) = \psi(t)/g(Z).
\label{eq:Zdot}
\end{equation}
The derived age-metallicity relation $Z(t)$ is depicted in Figure \ref{fig:Z}. In broad agreement with previous results, it exhibits a rapid increase in metallicity at early epochs with an enrichment timescale of about 4 Gyr, followed by a slow evolution around solar values at ages between 4 and 10 Gyr and an upward trend toward supersolar metallicities at more recent times 
\citep[e.g.,][]{Haywood2013,Snaith2015,Sharma2021}.

We note in passing
that the age-metallicity distribution may actually consist of two distinct populations, an old and a younger sequence   corresponding to the formation of, respectively, the thick and the thin disk \citep{Nissen2020}. An analysis of the implications of this scenario is postponed to future work.

\subsection{Planet Occurrence Frequency Around FGK Stars}

Exoplanet statistics in the inner regions of FGK dwarf stars have been investigated using the large and homogeneous sample from the {Kepler} mission \citep{Thompson2018}. {Kepler} planets commonly reside in multiplanet systems, and the integrated occurrence rate (the average number of planets per star) for exoplanets with radii in the range $1-20\, R_\oplus$ and orbital periods up to $P=400\,$days is \citep{Zhu2021}

\begin{equation}
{\eta}_{P}=1.23\pm 0.06.
\end{equation}
Earth-size exoplanets in Earth-like orbits are not well probed by {Kepler}, and estimates of their frequencies are more uncertain. A recent analysis by \citet[][see also \citealt{Burke2015}]{Bryson2020} yields
\begin{equation}
{\eta}_{1}=0.015^{+0.011}_{-0.0007}
\end{equation}
for the occurrence rate around GK dwarf stars
of terrestrial planets within 20\% of Earth’s orbital period and radius. 

The planet radius-orbital period parameter space defining $\eta_{1}$
is a subset of the larger parameter space for $\eta_\oplus$, the occurrence rate of Earth-size rocky planets in the HZ (hereafter ``temperate terrestrial planets" or TTPs for short), roughly defined as the region around a Sun-like star in which a rocky planet with an Earth-like atmospheric composition can sustain liquid water on its surface \citep{Kasting1993}.  The basic requirement for surface liquid water is predicated on a subset of the minimum conditions needed for a simple, microbial biosphere. Defining $\eta_\oplus$ as the occurrence rate of TTPs
with radii between 0.5 and 1.5 $R_\oplus$ and orbiting stars with effective temperatures between 4800 and 6300 K, \citet{Bryson2021} recently derived 
\begin{equation}
0.37^{+0.48}_{-0.21}<\eta_\oplus< 0.60^{+0.90}_{-0.36},
\end{equation}
where the errors reflect 68\% confidence intervals and the lower and upper bounds correspond to different
completeness corrections. This occurrence rate uses the conservative HZ estimates from \citet{Kopparapu2014}. 

Below, we shall adopt a fiducial present-day occurrence rate  of $\eta_\oplus=0.24$. This is the standard value used in forecasting TTP yields from direct-imaging future flagship missions like HabEx and LUVOIR, and is based on the NASA ExoPAG SAG13 meta-analysis of {Kepler} data \citep{Kopparapu2018}. Note that, in the language of Drake's equation (Equation (\ref{eq:Drake})), $\eta_\oplus \equiv f_pn_e$.

For comparison, the frequency of giant gaseous  planets with radii $>4\,R_\oplus$ and orbital periods $P<400\,$days is estimated by \citet{Zhu2021} to be
\begin{equation}
\eta_{\rm GP}=0.16\pm 0.015.
\end{equation}

\subsection{Dependence of Planet Frequency on Stellar Metallicity}

In the context of core-accretion planet formation theory, metal-rich protoplanetary disks have enhanced surface densities of solids, leading to the more efficient formation of the rocky cores of gas giant planets \citep{Pollack1996,Ida2004}. Is is well established observationally that metal-rich stars are more likely to host close-in giant planets \citep[e.g.,][]{Fisher2005,Petigura2018,Zhu2019},
with an occurrence rate enhancement as a function of metallicity of the form

\begin{equation}
f(Z)={\cal A}(Z/Z_\odot)^{2}.
\end{equation}
For a sample of stars with metallicity distribution $g(Z)$, the normalization constant ${\cal A}$ above is related to the integrated frequency of giant planets by \citet{Zhu2016}

\begin{equation}
\eta_{\rm GP}=\int g(Z)f(Z)dZ.
\end{equation}
With the adopted MDF (Fig. \ref{fig:MDF}), one derives ${\cal A}=0.1369$.

Small planets show a weaker frequency-metallicity correlation \citep{Sousa2008,Buchhave2012,Petigura2018}. Most recently, \citet{Lu2020} were unable to confirm or reject a relationship between planet occurrence and host star metallicity for rocky planets with radii $\lta 2R_\oplus$. 
In line with these analyses, no frequency-metallicity correlation for terrestrial planets will be assumed in our  treatment. We shall also ignore possible correlations between small planet occurrence rates and $\alpha$-element enhancement \citep{Bashi2020}. 

\subsection{Critical Metallicity for Terrestrial Planet Formation}

In the core-accretion model of planet formation heavy elements are necessary to form the dust grains and planetesimals that build planetary cores.
\citet{Johnson2012} estimated a minimum  metallicity $Z_c$ for planet formation by comparing the timescale for dust grain growth and settling to that for protoplanetary disk photoevaporation. They found that,  for an Earth-size planet to form, a disk of surface density $\Sigma(r)$ must have a metallicity

\begin{equation}
Z\gta 0.1\,Z_\odot \left[{\Sigma(1\,{\rm AU})
\over 10^3\,{\rm g\,cm^{-2}}}\right]^{-0.48}.
\label{eq:Mcrit}
\end{equation}
Given the observed MDF (Fig. \ref{fig:MDF}), the existence of a fiducial metallicity floor $Z_c = 0.1\,Z_\odot$ for the formation of terrestrial planets will only impact a small fraction of the census population in the solar neighborhood. Nevertheless, in our equations below we shall formally multiply the integrated occurrence rate of TTPs by the Heaviside function $\theta(Z$--$Z_c)$.

\begin{figure}[h!]
\centering
\vspace{+0.cm}
\includegraphics[width=\hsize]{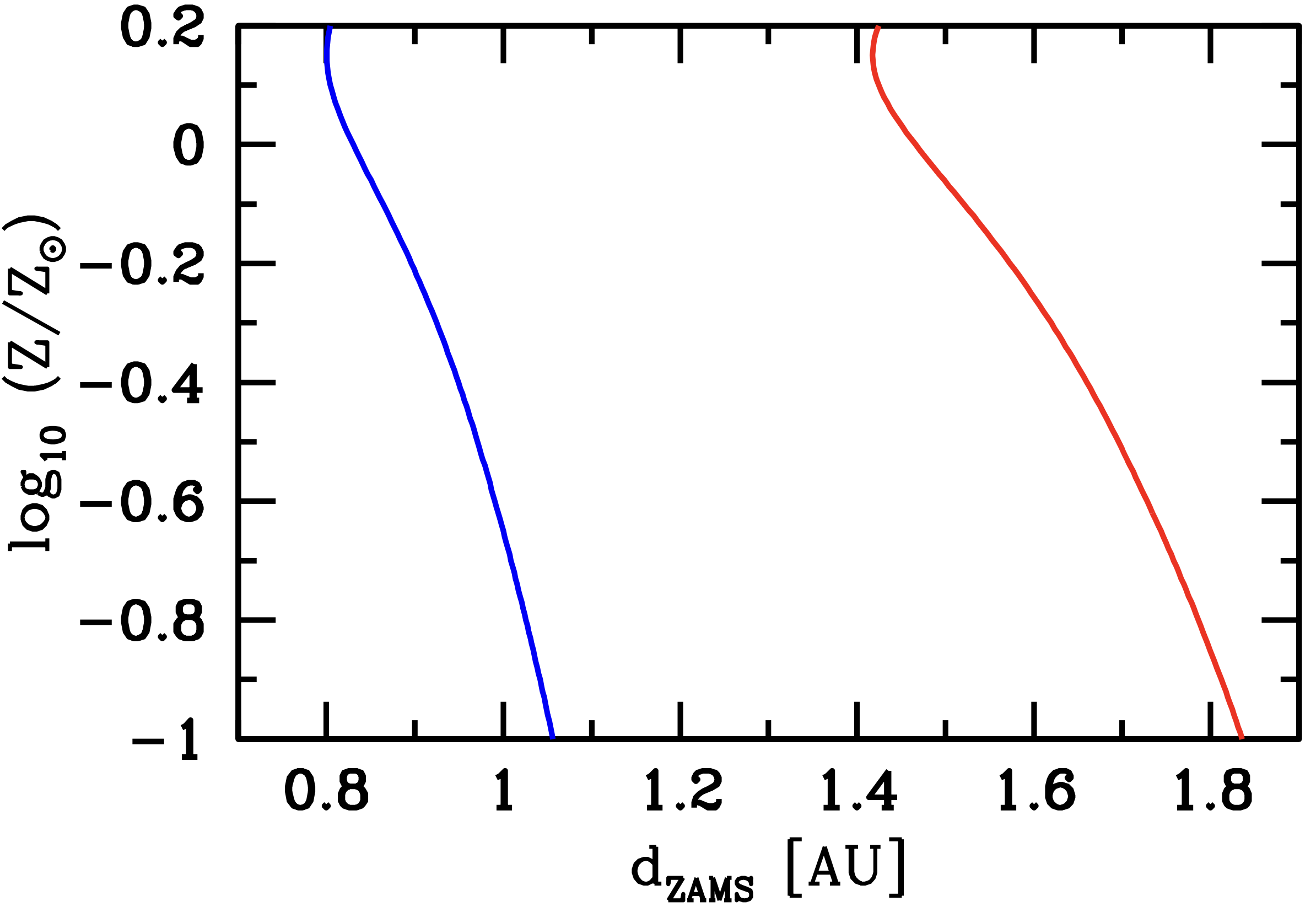}
\caption{ZAMS HZ limits for a $1\,M_\oplus$ 
planet around $m=1$ G-type host stars of different metallicities. The inner HZ (blue curve) is defined by the runaway greenhouse limit, while the red curve marks the outer HZ -- the maximum greenhouse limit. Stellar effective temperatures and luminosities were computed following \citet{Tout1996}, 
and the $y$-axis covers the range from $(L_{\rm ZAMS}/L_\odot, T_{\rm eff})=(0.70, 5635\,{\rm K})$ at $Z=1.6\,Z_\odot$ to $(L_{\rm ZAMS}/L_\odot, T_{\rm eff})=(1.34, 6464\,{\rm K})$ at $Z=0.1\,Z_\odot$. The effective incident fluxes determining the inner and outer HZ were estimated using the parametric formulae of \citet{Kopparapu2014}.
}    
\label{fig:HZ}
\end{figure}

\subsection{Effect of Metallicity on the HZ}

Low-metallicity stars have higher luminosities $L_{\rm ZAMS}$ and higher effective temperatures $T_{\rm eff}$ than metal-rich stars of the same mass \citep[e.g.,][]{Tout1996}. An $m=1$ zero-age main-sequence (ZAMS) star of metallicity $Z=0.1\,Z_\odot$, for example, is 80\% brighter, 13\% hotter, and has a larger HZ than its solar-metallicity counterpart. Conversely, an F-type star with $m=1.5$ and $Z=0.1\,Z_\odot$  has a main-sequence lifetime of only 1.8 Gyr (vs. 2.7 Gyr at solar metallicity), and is unlikely to be a good candidate for harboring continuously habitable planets.

We assess the impact of stellar metallicity on the habitability of terrestrial exoplanets around GK stars (see also \citealt{Danchi2013,Valle2014,Truitt2015})
adopting the conservative HZ estimates of \citet{Kopparapu2014}, where the inner and outer edges of the continuously HZ are defined by the ``runaway greenhouse" (atmosphere becomes opaque to outgoing thermal radiation owing to excess amounts of H$_2$O)  and ``maximum greenhouse” (increased reflectivity of a thick CO$_2$ atmosphere wins out over greenhouse effect) limits.
The 1D, radiative–convective, cloud-free climate models of \citet{Kopparapu2013} provide critical  values for the effective flux $S_{\rm eff}$ -- the normalized value of solar constant required to maintain a given surface temperature -- as a function of the effective temperature of the host star

\begin{equation}
S_{\rm eff}=S_{\rm eff,\odot}+aT_\star+bT_\star^2+cT_\star^3+dT_\star^4,
\label{eq:Seff}
\end{equation}
where $T_\star=T_{\rm eff}-5760\,$K and the coefficients $(a,b,c,$ and $d)$ for the runaway greenhouse and maximum greenhouse limits are listed in \citet{Kopparapu2014}. Stellar effective temperatures and luminosities on the ZAMS were computed as a function of metallicity following \citet{Tout1996}.
The corresponding HZ distances can be calculated using the relation 

\begin{equation}
d_{\rm ZAMS}=\left({L_{\rm ZAMS}/L_\odot\over S_{\rm eff}}\right)^{0.5}\,{\rm AU},
\label{eq:dHZ}
\end{equation}
where $L_{\rm ZAMS}/L_{\odot}$ is the luminosity of the star in solar units.\footnote{Throughout this paper, the $\odot$ subscript denotes present-day values.}~
Figure \ref{fig:HZ} depicts the predicted variations in HZ boundaries as a function of stellar metallicity for an $m=1$ star. The effective stellar fluxes at the inner and outer edges of the HZ increase at low $Z$s  because the calculated planetary albedos become higher as the star’s radiation is shifted toward the blue, a dependence that is stronger at the outer HZ boundary because of the importance of Rayleigh scattering in dense CO$_2$ atmospheres \citep{Kasting2014}. Note, e.g., how a $1\,M_\oplus$ planet at 1 au from an  $m=1$ ZAMS G-class dwarf of metallicity $0.1\,Z_\odot$ is just outside the inner edge of its host's HZ, as this is evaluated at $1.05-1.83$ au.

Figure \ref{fig:HZ} shows how, at low metallicities, the HZ moves further away from the host star and is about 25\% wider. These differences in HZ boundaries may be relevant for present and upcoming planet-finding surveys around low-metallicity stars \citep{Dedrick2020}. Because old age correlates with low stellar metallicity, one might expect $\eta_\oplus$ to be larger at earlier times if other factors, like  the efficiency of planetary formation and planetary spacing, were equal. Given the relatively flat period distribution, $df/d\ln p\propto p^{0.2}$, observed for long-period {Kepler} planets \citep[e.g.,][]{Bryson2020}, however, the impact of such a shift in HZ boundaries on the predicted occurrence rates of TTPs around GK dwarfs in the young Galaxy should be rather minor. 

\subsection{HZ Lifetime}

The boundaries of a radiative HZ are not temporally or spatially static but ``migrate" outward over the course of a star's main-sequence phase. The secular increase in stellar luminosity results in a runaway greenhouse event, which, in the case of Earth, will cause the cessation of habitable conditions and the likely extinction of our biosphere about 1.5-2 Gyr in the future \citep[e.g.,][]{Goldblatt2012,Sousa2020}, well before the Sun becomes a red giant. The transitory nature of the residence of a TTP within an HZ has strong astrobiological implications and
can be described by the time a planet, located at a given distance from a star, spends within the HZ \citep{Danchi2013,Rushby2013,Waltham2017}. Here, we have used the formulae and best-fitting coefficients of \citet{Hurley2000} for the time-dependent luminosities and radii of stars on the main sequence to derive estimates for the change in the HZ boundaries over time.

\begin{figure}[!h]
\centering
\vspace{0.cm}
\includegraphics[width=\hsize]{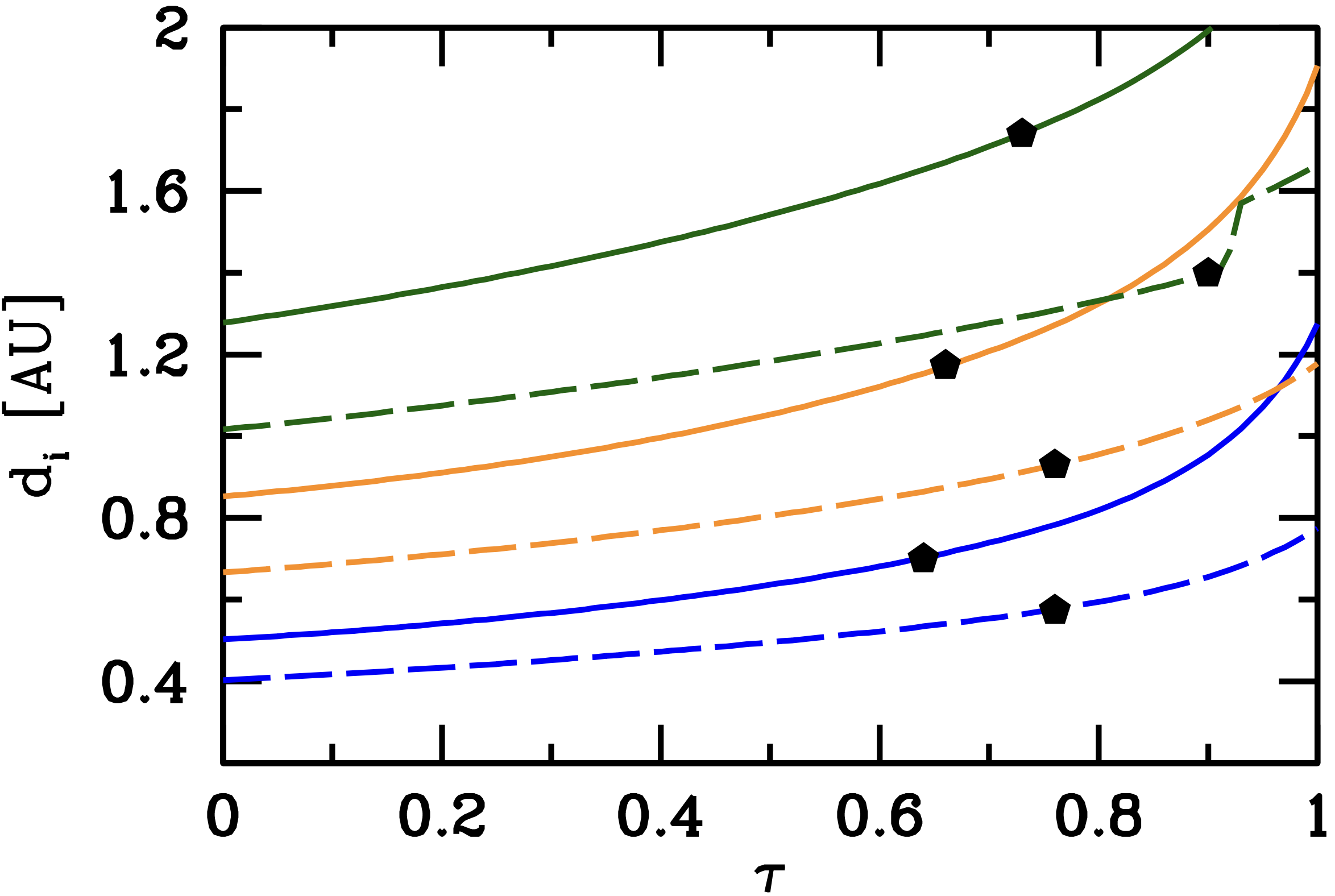}
\hspace*{0.1cm}
\includegraphics[width=\hsize]{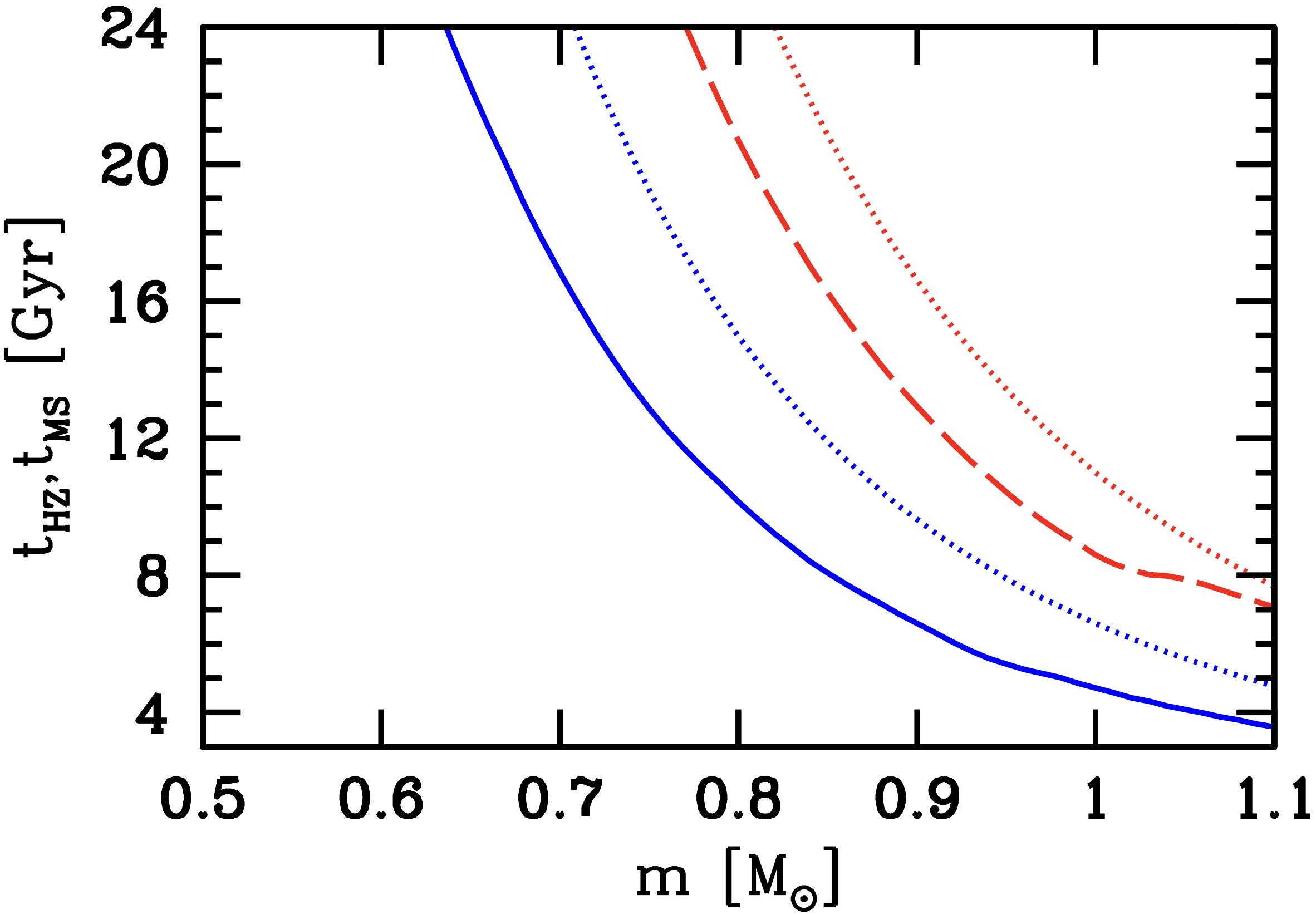}
\caption{Time-dependent HZ boundaries around GK dwarf stars. {Top panel:} evolution of the inner
boundary of the HZ, $d_i$ (in au), from the ZAMS (corresponding to $\tau=0$) to the  end of the main sequence, as a function of $\tau=t/\tMS$ for stars of different mass and metallicities. Solid lines: $m=0.7$ (blue), $m=0.9$ (orange), and $m=1.1$ (green), all calculated at $Z=0.1\,Z_\odot$. Dashed lines: same for $Z=Z_\odot$. The point on each curve denotes the time when a hypothetical planet, formed in the center of the HZ at the ZAMS stage, becomes uninhabitable. {Bottom panel:} HZ lifetime $\tHZ$ (in Gyr) as a function of  stellar mass and metallicity. Solid blue line: $Z=0.1\,Z_\odot$. Dashed red line: $Z=Z_\odot$. The dotted lines show the corresponding main-sequence timescale.
}
\label{fig:THZ}
\end{figure}

Figure \ref{fig:THZ} (top panel) shows the evolution of the inner edge of the HZ, computed as before from the prescriptions of \citet{Kopparapu2014}, as a function of time $\tau=t/\tMS$ for GK dwarf stars of different masses and metallicities. The point on each curve marks the characteristic ``HZ lifetime," $\tHZ(m,Z)$, when a hypothetical planet formed {\it in the center} of the HZ at the ZAMS stage enters into the hot zone of the host star, undergoes a runaway greenhouse event, and becomes uninhabitable \citep{Rushby2013}.\footnote{Note that 
the habitable lifetime in fact changes with star-planet
separation, gradually increasing  between the inner and outer edges of the HZ, and that the full distribution 
of $\tHZ$ with distance should be taken into account in more advanced modeling \citep[e.g.][]{Waltham2017}.}\,
Lower-mass stars, while characterized by longer main-sequence lifetimes, have proportionally smaller $\tHZ/\tMS$ ratios than higher-mass stars, the result of their lower rates of stellar luminosity evolution \citep{Rushby2013}. At fixed mass, main-sequence lifetimes are shorter and the total luminosity change over the main sequence is larger for lower-metallicity stars \citep{Truitt2015}. In the bottom panel of Figure \ref{fig:THZ} we compare the ``HZ lifetime" with the main-sequence timescale
as a function of  stellar mass and metallicity. Over the plotted mass interval, the ratio $\tHZ/\tMS$ ranges from 0.75 to 0.90 at solar metallicities, and from 0.65 to 0.75 at $0.1\,Z_\odot$. In absolute value terms, we find HZ lifetimes that are longer than the age of the Galaxy for $m<0.9$ ($Z=Z_\odot$) and $m<0.75$ ($Z=0.1\,Z_\odot$).

Below, we shall assume that the evolution of main-sequence stars -- rather than biogeochemical 
processes -- is the only factor controlling the collapse of the HZ zone and the reduction of the 
biosphere lifespan. As the HZ expands outward due to the effects of stellar evolution, any planets that were initially beyond the boundaries of the HZ -- so-called ``cold start" icy planets -- could potentially become habitable at later times as the HZ reaches them. We shall neglect this possibility 
below as the delayed habitability of such globally glaciated exoplanets remains dubious \citep[e.g.,][]{Yang2017}.

\section{Exoplanets Around K Dwarfs}

We can now cast our model for the time evolution of the exoplanet population in the solar neighborhood into a set of 
rate equations that can be then be integrated as a function of time. Let us focus on K-class dwarfs with masses between $m_1=0.45$ and $m_2=0.80$ and main-sequence and HZ lifetimes that are typically longer than the age of the Galaxy at the metallicities of interest here. K stars may be better candidates in the search for biosignatures than G dwarfs, as they are more abundant, evolve less quickly on the main sequence, and provide their planets a stable HZ \citep[e.g.,][]{Cuntz2016,Tuchow2020}. They also offer a longer photochemical lifetime of methane in the presence of oxygen compared to G dwarfs and, being dimmer, provide a better planet-star contrast ratio in direct-imaging observations \citep{Arney2019}.\footnote{
The photochemical lifetime of methane in oxygenated atmospheres is even longer around M dwarfs \citep{Segura2005}, but M dwarf planet habitability may be hindered by extreme stellar activity and a prolonged superluminous pre-main-sequence phase.}\,
For the assumed Kroupa IMF, the fraction of stars that are classified as K-type is

\begin{equation}
{\cal F}(m_1,m_2)\equiv \int_{0.45}^{0.80}
\phi(m)dm=0.14.
\label{eq:Ktype}
\end{equation}
Note that if the lower bound on habitability corresponded to the spectral type K5 instead ($m=0.65\,\msun$), the factor ${\cal F}$ in Equation (\ref{eq:Ktype}) would decrease by a factor of 3.5, while the inclusion of M dwarfs below 0.45 $\msun$ would boost the same integral by a factor of 6. Of course, for the occurrence rate of TTPs what matters is the product $\eta_\oplus
\times {\cal F}$, so one could group some of the uncertainties related to the lower habitability bound into the factor $\eta_\oplus$.

\subsection{Rate Equations}

We can now track the evolution of the mean number of  K-type stars in the solar neighborhood, $N_{K}(t)$, and the abundance of giant planets and TTPs around them, $N_{\rm GP}(t)$ and $N_\oplus(t)$, by numerically integrating over time the corresponding rates
\begin{equation}
\begin{aligned}
\dot N_K (t) & =  N_\star(t_0) \psi(t) {\cal F},\\
\dot N_{\rm GP}(t) & =  f[Z(t)] \dot N_K(t),\\
\dot N_{\oplus}(t) & = \eta_\oplus \theta[Z(t)-Z_c] \dot N_K(t).
\label{eq:Rates}
\end{aligned}
\end{equation}
Here, in the equation for $\dot N_{\oplus}(t)$, we have assumed that exoplanets enter the HZ immediately after formation, and interpreted the parameter $\eta_\oplus$ as the occurrence rate of TTPs around K-class stars on the ZAMS. \footnote{For more massive FG-type stars with main-sequence and HZ lifetimes that are shorter than the age of the Galaxy, the rate equations
(\ref{eq:Rates}) take the more complicated form
\begin{equation}
\begin{aligned}
\dot N_{\rm FG} (t) & = 
N_\star(t_0) 
\phi(m) [\psi(t) - \psi(t-\tMS)],\\
\dot N_{\rm GP}(t) & =  f[Z(t)]\,\dot N_{FG}(t),\\
\dot N_{\oplus}(t) & = \eta_\oplus N_\star(t_0)\phi(m)\,\theta[Z(t) -Z_c]\psi(t)\\
& - \eta_\oplus N_\star(t_0)\phi(m)\,\theta[Z(t-\tHZ)-Z_c]\psi(t-\tHZ).
\label{eq:GFRates}
\end{aligned}
\end{equation}
The second term in the last equation  
approximately corrects the rate of newly formed TTPs for the amount that has ``migrated" out of the HZ over the course of the star’s main-sequence lifetime. Note that, because of the mass dependence of $\tHZ\le \tMS$, the rate equations must now be integrated in bins of stellar mass.}\, 
The equations above must  be supplemented by Equation (\ref{eq:Zdot}) for the evolution of the stellar  metallicity $Z(t)$. 

\begin{figure}[h!]
\centering
\vspace{+0.cm}
\includegraphics[width=\hsize, 
trim={0  2.3cm 0 0},clip]{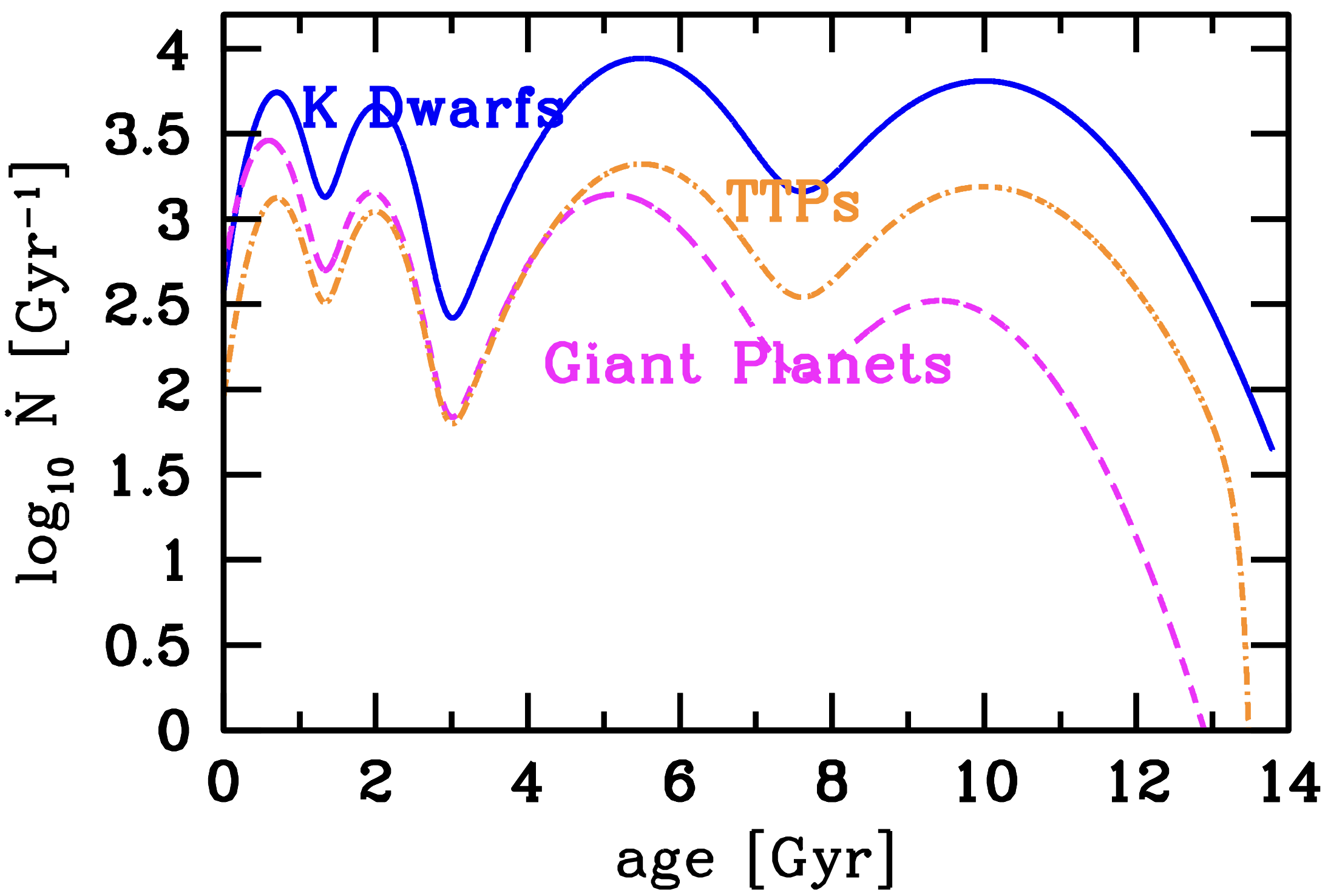}
\includegraphics[width=\hsize]{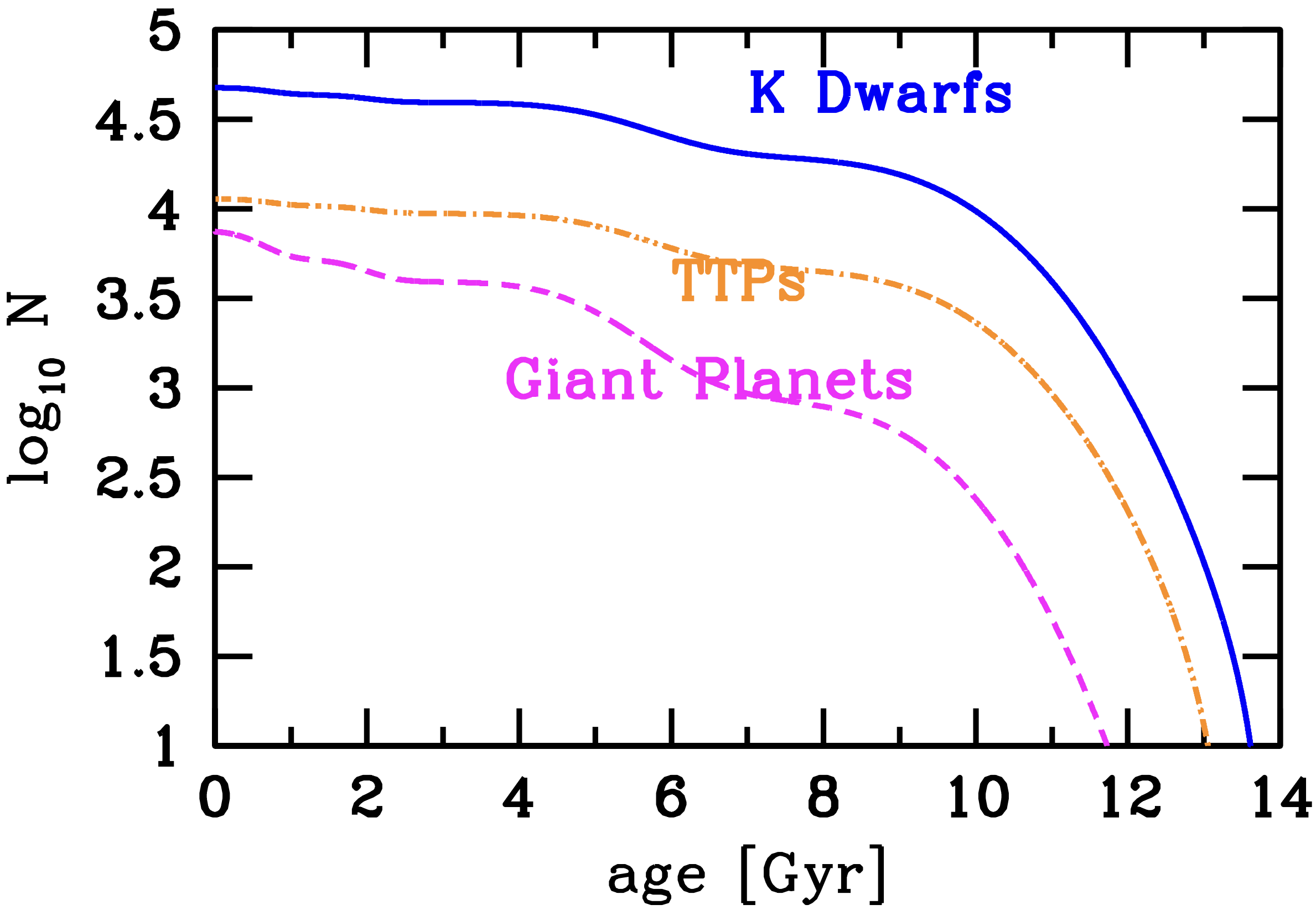}
\caption{Exoplanet formation history in the solar neighborhood.
{Top panel:} K-class dwarf (solid curve), giant planet (dashed curve), and TTP (dotted-dashed curve) formation rates in a ``solar vicinity" sphere of 100 pc radius, as a function of age $t_0-t$. These estimates assume $\eta_{\rm GP}=0.16$ and $\eta_\oplus=0.24$ (see text for details).
{Bottom panel:} cumulative number counts resulting from the integration of Equation (\ref{eq:Rates}). Note how the solar system is younger than 77\% of all TTPs, and has an age that is comparable to that of the median giant planet.
}    
\label{fig:Counts}
\end{figure}

Figure \ref{fig:Counts} shows the episodic formation history of exoplanets that results from the integration of Equations (\ref{eq:Rates}) for $\eta_{\rm GP}=0.16$ and $\eta_\oplus=0.24$. In our ``solar vicinity" sphere of 100 pc radius, there are currently about $11,000\,(\eta_\oplus/0.24)$  TTPs around K dwarfs. They have a  median age of 6.2 Gyr and 77\% of them are older than the solar system. The minimum metallicity threshold for Earth-size planet formation of \citet{Johnson2012} does not significantly affect these numbers as the vast majority of star formation has taken place at $Z>0.1\,Z_\odot$. By contrast, the $f(Z)$ modulation of the giant planet occurrence rate results in later typical formation times and shifts their median age to 3.9 Gyr, with terrestrial planets vastly outnumbering giant planets at early times.

The early formation of TTPs in the solar vicinity occurred largely in two major episodes of enhanced star formation, starting with the emergence of the thick disk about 11 Gyr ago and followed by a second event that lasted 3.5 Gyr, peaked 5.5 Gyr ago, and involved more than 40\% of the total stellar counts today. The five planet system Kepler-444, orbiting a metal-poor $11.2\pm 1.0\,$ Gyr old star, shows that thick-disk stars were indeed the hosts of some of the oldest terrestrial planets \citep{Campante2015}. The duration and size of the second major star formation surge suggest an external agent, perhaps a  merger with a gas-rich satellite galaxy \citep{Mor2019} or the first passage of the Sagittarius dwarf galaxy through the Milky Way's disk \citep{Ruiz-Lara2020}. Consistently with the Principle of Mediocrity, the solar system formed near the peak of this second episode. Over the last 4 Gyr the abundance of TTPs around K stars has increased by +24\%, while that of giant planets has actually doubled.

\section{Simple Life in the Solar Neighborhood}

In the coming decades, advanced space- and ground-based observatories will allow an unprecedented opportunity to probe the atmospheres and surfaces of TTPs in search for signs of life or habitability.
The discovery of extraterrestrial life would be a landmark moment in the history of science, with implications that would reverberate throughout all of society. 
Much of the early history of life on Earth has been dominated by methanogenic microorganisms, and methane in anoxic, Archean-like atmospheres
is one of the most promising exoplanet spectroscopic biosignatures \citep{Kaltenegger2017,Schwieterman2018,Thompson2022}. 
In spite of the recent swift developments in  astrophysics and planetary sciences described in the previous sections, however, the probability of abiogenesis on Earth-like planets is currently unknown, as unknown are the characteristic timescales over which biochemical complexity actually evolves. The rapid emergence of life in the history of Earth 
%or as far back as 4.1 Gyr ago according to more disputed evidence of $^{13}$C-depleted zircon deposits \citep{Bell2015} 
has been used to argue for a high abiogenesis rate \citep[e.g.,][]{Lineweaver2002,Kipping2020,Whitmire2022}. A Bayesian framework may naturally account for the anthropic bias that, if the timescale for intelligence evolution is long, life’s early start may simply be a prerequisite to our existence, rather than evidence for simple primitive life being common on Earth-like worlds \citep{Spiegel2012}. By means of  this methodology, the recent analysis by \citet{Kipping2020} shows that a fast abiogenesis scenario is at least three times more likely than a slow one.

In preparation for the next generation of space- and ground-based instruments, it seems interesting to conjecture on today's prevalence and time-varying incidence of  microbial life-harboring worlds in the solar neighborhood under the hypothesis of a rapid abiogenesis process. We follow previous work \citep{Spiegel2012,Scharf2016,Chen2018,Kipping2020} and describe abiogenesis as a stochastic Poisson process defined by a (uniform) rate parameter $\lambda_\ell$ -- the mean number of abiogenesis events occurring per Earth-like planet in a fixed time span, which we set equal to 1 Gyr. The probability of achieving at least one successful abiogenesis event over a time interval $t-t'$ since a given planet first became habitable at $t'$ is then given by
\begin{equation}
P_\ell(t-t')= 1-e^{-\lambda_\ell\,(t-t')}.
\label{eq:Plife}
\end{equation}
The time-dependent probability that life emerges on a TTP can then be expressed as the product $P_\ell(t-t')P({\ell|\rm HZ})$, i.e. we distinguish here between the population of planets that are ``temperate"
(i.e. Earth-size rocky planets in the continuously HZ) and the subset that are actually ``habitable," i.e. ``Earth-like" in a more detailed biochemical and geophysical sense, and where simple life will eventually arise. The number of these ``Earth analogs" that formed (and became habitable) between time $t'$ and $t'+dt'$ and where life emerged by time $t$ is $dN_\ell(t)=P({\ell|\rm HZ})P_\ell(t-t') \dot N_\oplus(t')dt'$. 
Assuming a probability $P({\ell|\rm HZ})$ that is independent of time, 
we can then write the mean number of life-hosting worlds present at time $t$ as the convolution integral

\begin{equation}
N_\ell(t)=P({\ell|\rm HZ}) \int_0^t \dot N_\oplus (t')\,
\left[1-e^{-\lambda_\ell(t-t')}\right]dt'.
\label{eq:Nlife}
\end{equation}
Their formation rate must be given by the time derivative of Equation (\ref{eq:Nlife}), yielding

\begin{equation}
\dot N_\ell(t)=P({\ell|\rm HZ}) \int_0^t \dot N_\oplus (t')\,
\lambda_\ell\,e^{-\lambda_\ell(t-t')}dt'. 
\label{eq:dNlife}
\end{equation}
Note that, in the formalism of Drake's Equation (\ref{eq:Drake}), $P({\ell|\rm HZ})=f_\ell$ in the limit of fast abiogenesis. 

The oldest, generally accepted evidence for life on Earth comes from observations of microbial sulfate reduction in the 3.48 Gyr Dresser Formation \citep[e.g.,][]{Lepot2020}. The Earth formed $4.54\pm 0.05$ Gyr ago \citep{Dalrymple2001},  and mineralogical evidence from detrital zircons indicates that liquid oceans must have been present on Earth’s surface $4.404\pm 0.008$ Gy ago \citep{Wilde2001}.
The maximum plausible value for the time interval over which at least one successful abiogenesis event occurred on Earth is therefore $\simeq  0.9-1\,$Gyr. This is conservatively long compared to the maximum-likelihood timescale for life to first appear after conditions became habitable, $\sim 190$ Myr,  inferred by \citet{Kipping2020}, and much larger than the uncertainty as to when Earth became suitable for life, justifying our approximation of starting the ``habitability clock" at formation.

\begin{figure}[t]
\centering
\vspace{+0.cm}
\includegraphics[width=\hsize]{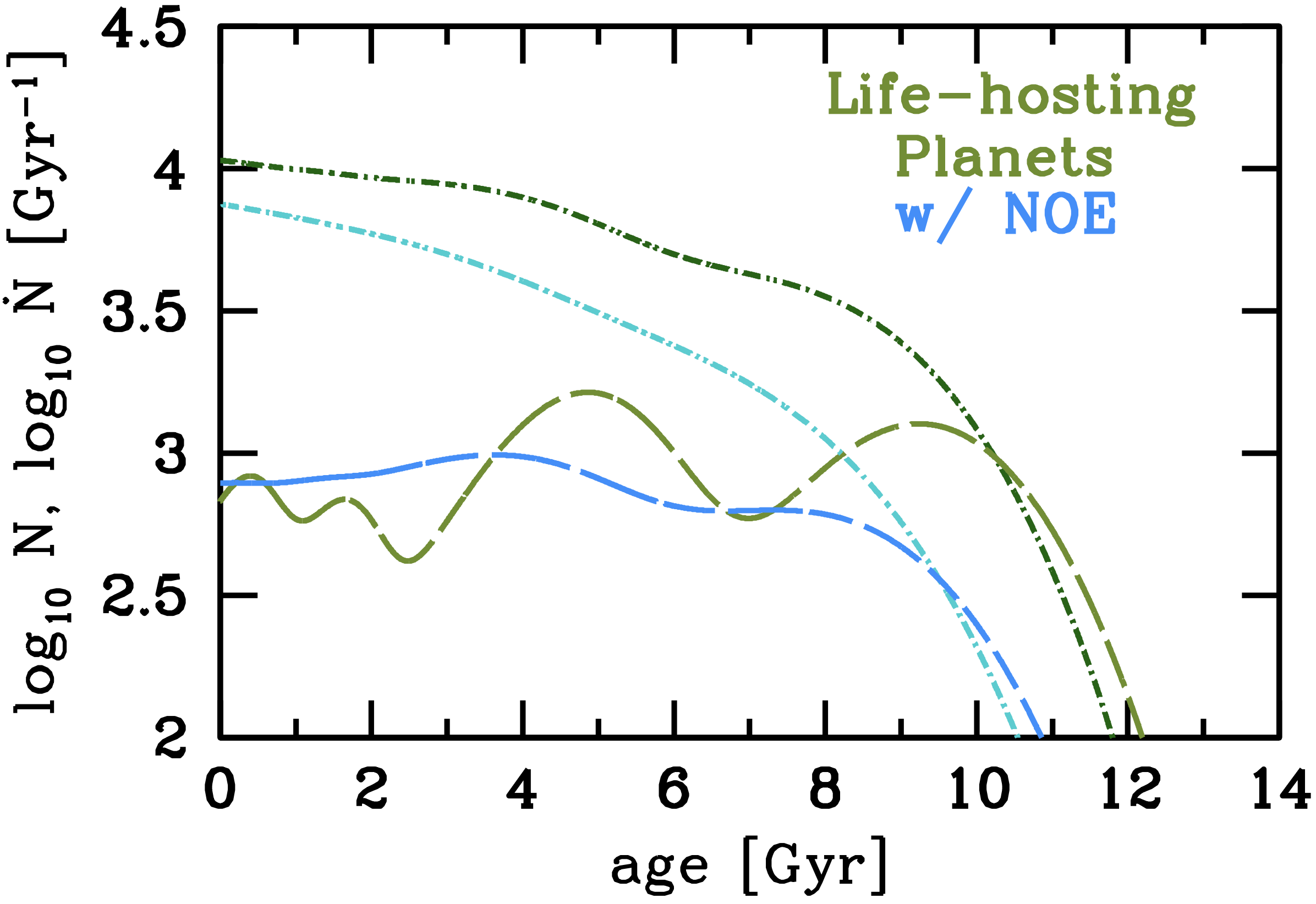}
\caption{Abiogenesis in the solar neighborhood. {Long-dashed green curve:} formation rate, $\dot N_\ell $, of life-bearing exoplanets as a function of age $t_0-t$. The calculation (Equation (\ref{eq:dNlife})) assumes a rapid abiogenesis process with rate parameter $\lambda_\ell=1\,$Gyr$^{-1}$, and  simple life eventually arising in all TTPs, $P({\ell|\rm HZ})=1$. 
{Dot-dashed green curve:} cumulative number counts of life-hosting exoplanets, $N_\ell$, resulting from the integration of Eq. (\ref{eq:Nlife}).
{Blue curves:} same for life-hosting planets undergoing 
a ``Neoproterozoic oxygenation event" (NOE) with 
rate parameter $\lambda_O^{-1}=3.9\,$Gyr.
}    
\label{fig:LIFE}
\end{figure}

We have integrated Equations (\ref{eq:Nlife}) and (\ref{eq:dNlife}) above assuming $\lambda_\ell=1\,$Gyr$^{-1}$, and plotted in Figure \ref{fig:LIFE} the resulting differential and cumulative number counts, $\dot N_\ell$ and $N_\ell$. For illustration, we have assumed in the figure a conversion factor $P({\ell|\rm HZ})=1$ between TTPs and life-hosting Earths. Naturally, life would be abundant (again, we are concerned here with the appearance of the earliest life forms, not of intelligent life) in the solar vicinity if abiogenesis was fast and early Earth-like conditions 
existed and were relatively common on other worlds for 1 Gy or more. The closest life-harboring exoplanet would be only 20 pc away if simple life arose as soon as it did on Earth in just 1\% of TTPs around K stars. Conversely, Earth would be the only life-hosting planet in the solar neighborhood if abiogenesis was successful in about 1-in-10,000 TTPs. If microbial life is abundant it is also old, as it would have emerged more than 8 Gyr ago in about one-third of all life-bearing planets today. Note how the convolution integral in Equation (\ref{eq:dNlife}) tends to smooth out the oscillations in $\dot N_\ell$ compared to the star and planet formation rates depicted in Figure \ref{fig:Counts}, and that the assumed abiogenesis characteristic timescale of $1/\lambda_\ell=1\,$Gyr shifts the median age of 
the $\sim 10,000\,P({\ell|\rm HZ})$ extrasolar biospheres predicted 
in the solar neighborhood to 5.7 Gyr.

\subsection{The Emergence of Oxygenic Atmospheres}

A critical issue in the search for extraterrestrial life is whether Earth-like conditions lead to ecosystems that progressively oxygenate their planet atmospheres roughly following Earth's oxygenation history.
The first oxygenation of Earth's atmosphere due to the emergence of photosynthesizing cyanobacteria happened about halfway through Earth’s history \citep{Luo2016}, but O$_2$ rose irreversibly to near present atmospheric levels only about between 800 and 550 Myr ago, during the NOE \citep{Och2012,Lyons2021}, an event accompanied by major biological upgrades. While the precise timing of the NOE remains subject of debate, 
our findings inevitably invite the question of whether and how often, given an habitable environment and following a successful abiogenesis event, the conditions for the beginnings of complex life may have arisen on exoplanets in the solar neighborhood. We can then consider a second stochastic process, labeled ``$O$" for ``oxygenation" and defined by a rate parameter $\lambda_O$, which can proceed only once abiogenesis (“$\ell$”) is successful. The inverse of $\lambda_O$ is the characteristic timescale it takes for the earliest forms of life to evolve and produce an NOE-like juncture. Consider again an Earth-like planet that formed at time $t'$. The joint probability density that abiogenesis was first successful at time $t''$ and was followed by an oxygenation event at time $t$ is then given by

\begin{equation}
p_O(t''-t',t-t'')=
\lambda_\ell\lambda_O\,e^{-\lambda_\ell (t''-t') 
-\lambda_O (t-t'')}.
\label{eq:Pjoint}
\end{equation}
The formation rate of simple life-hosting planets undergoing an NOE can then be written as  
\begin{equation}
\dot N_O(t)=P({\ell|\rm HZ}) \int_0^t dt''
\int_0^{t''} dt' 
\dot N_\oplus (t')p_O.
\label{eq:dNO}
\end{equation}
We have integrated this equation assuming $\lambda_O^{-1}=
3.9\,$Gyr and plotted the results in Figure \ref{fig:LIFE}. Because of the considerable delay between planet formation and NOE, the $\sim 7500\,P({\ell|\rm HZ})$ worlds with strong oxygenic atmospheric biosignatures predicted to exist in the solar neighborhood have a formation rate that peaked 5 Gyr ago and a median age comparable to that of the solar system. 

\section{Summary and Discussion}

The search for habitable exoplanets and extraterrestrial life beyond Earth is one of the greatest scientific endeavors of all time. The high frequency of terrestrial planets in the HZs around dwarf stars implied by {Kepler} observations makes it timely to develop and explore new tools -- beyond the probabilistic Drake equation -- for statistical exoplanet population and astrobiology studies that may help directfuture mission designs and observational efforts. In particular, one would like to understand -- given a model of habitability and biosignature genesis -- the formation history of simple life-harboring environments in the local volume, and identify how potential atmospheric biosignature yields change as a function of stellar properties like age, mass, and metallicity.

The approach we have described in this work is based on a system of simple ordinary differential equations, rewritten below for the convenience of the reader

\begin{equation}
\begin{aligned}
\dot N_K(t) & =  N_\star(t_0) \psi(t) {\cal F},\\
\dot Z(t) &= \psi(t)/g(Z),\\
\dot N_{\rm GP}(t)& =  f(Z) \dot N_K,\\
\dot N_{\oplus}(t) & = \eta_\oplus \theta(Z-Z_c) \dot N_K,\\
\dot N_\ell(t) &= P({\ell|\rm HZ}) \int_0^t \dot N_\oplus (t')\,
\lambda_\ell\,e^{-\lambda_\ell(t-t')}dt',\\
\dot N_O(t) & =P({\ell|\rm HZ}) \int_0^t dt''
\int_0^{t''} dt' 
\dot N_\oplus (t')p_O,
\end{aligned}
\label{eq:ODEs}
\end{equation}
which track the evolution of the mean number $N_K$ of K-type stars in the solar neighborhood, their metallicity $Z=Z(t)$, and the abundances $N_{\rm GP}$, $N_{\oplus}$, $N_\ell$, and $N_O$ of giant planets, TTPs, life-harboring worlds, and 
planets with oxygen-rich atmospheres, respectively
These rate equations  provide a time-dependent mapping between star formation, metal enrichment, and the occurrence of potentially habitable exoplanets over the chemo-population history of the solar neighborhood, and presents a useful basis for testing hypotheses about Earth-like environments and life beyond the solar system. The new framework can be easily adapted to incorporate the hierarchy of astrophysical and biological processes that regulate the age-dependent inventory of any key planet population.

We have numerically integrated the equations above adopting the recent tally of nearby stars ($N_\star$) and white dwarfs from Gaia EDR3 \citep{Smart2021}, the episodic SFH
($\psi$) of the solar neighborhood  as reconstructed by \citet{Alzate2021} and \citet{Ruiz-Lara2020}, the MDF ($g$) from the GALAH$+$TGAS spectroscopic survey of dwarf stars in the solar galactic zone \citep{Buder2019}, and assuming an age-metallicity relation. In our model, the function $f(Z)$ describes the metallicity modulation of the occurrence rate
of giant gaseous planets (with integrated frequency today $\eta_{\rm GP}=0.16$, \citealt{Zhu2021}), 
and we take $\eta_\oplus=0.24$ for the fiducial  occurrence rate  of TTPs \citep{Kopparapu2018} around K-class stars on the ZAMS. There is a  minimum metallicity threshold for Earth-size planet formation of $Z_c=0.1\,Z_\odot$ \citep{Johnson2012}. 
Following earlier work \citep[e.g.][]{Spiegel2012}, we describe abiogenesis as a stochastic Poisson process defined by a (uniform) rate parameter $\lambda_\ell$, and denote with $P({\ell|\rm HZ})$ the (constant) probability that a seemingly potentially habitable planet in the HZ was at early times ``Earth-like" in a more detailed biochemical and geophysical sense and eventually became inhabited by life. A second stochastic process, an oxygenation event defined by a rate parameter $\lambda_O$,
can proceed only once abiogenesis is successful.

Our main results can be summarized as follows.

\begin{enumerate}
%[leftmargin=*]

\item The formation of exoplanets in the solar vicinity followed two major events of enhanced star formation, starting with the emergence of the thick disk about 11 Gyr ago and followed by a second event that peaked 5.5
Gyr ago, lasted 3.5 Gyr, and produced more than 40\% of all stars today. The solar system formed in the second star formation surge and was likely triggered by an external agent, perhaps a merger with a gas-rich satellite 
galaxy \citep{Mor2019}  or the first passage of the Sagittarius dwarf galaxy through the Milky Way’s disk
\citep{Ruiz-Lara2020}.

\item Within 100 pc from the Sun, there are as many as $11,000\,(\eta_\oplus/0.24)$ TTPs around  K-type stars. 
About 77\% of all TTPs in the solar neighborhood  are older than the solar system.

\item The metallicity modulation of the giant planet occurrence rate results in a later typical formation time, with TTPs vastly outnumbering giant planets at early times.
Over the last 4 Gyr, the abundance of TTPs around K stars has increased by only +24\%, while that of giant planets
has doubled. The existence of a fiducial metallicity floor for the formation of terrestrial
planets impacts only a small fraction of the census population in the solar neighborhood, as the vast majority of star formation has taken place at $Z>0.1\,Z_\odot$.

\item 
The closest life-harboring Earth analog would be less than 20 pc away if microbial life arose as soon as it did on Earth in $\gta 1$\% of the TTPs around K stars.  Conversely, Earth would be the only life-hosting planet in the solar neighborhood if abiogenesis was successful in about 1-in-10,000 TTPs. If simple life is abundant (fast abiogenesis with characteristic timescales $1/\lambda_\ell=1\,$Gyr), it is also old, as it 
would have emerged more than 8 Gyr ago in about one-third of all life-bearing planets today.

\end{enumerate}

We finally note that errors in the number counts of exoplanets are likely dominated by planet occurrence rates ($\eta_{\rm GP}$ and $\eta_\oplus$), which are uncertain at the $\sim 0.1-0.5$ dex level due to incompleteness in long-period candidates. Comparable systematic errors may be associated with uncertainties in the IMF, SFH, and metallicity effects. Needless to say, in the case of life-hosting worlds, the precise values of $P({\ell|\rm HZ})$, $\lambda_\ell$, and $\lambda_O$ are currently unknown and remain a matter of speculation. Our work says nothing about how difficult or easy abiogenesis really is, a question that must ultimately be answered empirically. Still, given a model of habitability and biosignature genesis, our approach  may provide a blueprint for assessing the prevalence of exoplanets and microbial life-harboring worlds over the chemo-population history of the solar neighborhood, gaining a sense of the atmospheric biosignature yields among potential target stars of different masses, ages, and metallicities, and guiding future observational efforts and experiments.

\section*{Acknowledgements}

Support for this work was provided by NASA through grant 80NSSC21K027. We acknowledge useful discussions on this project with J. Alzate, M. Bisazza, F. Haardt, D. Lin, and R. Murray-Clay, and the hospitality of New York University Abu Dhabi during the completion of this study. The author would also like to thank the referee for a number of constructive comments and suggestions that greatly improved the paper.

\bibliographystyle{apj}
\bibliography{paper}

\label{lastpage}
\end{document}